\documentclass[onecolumn]{pasj00}
\begin{document}
\SetRunningHead{K. Takahashi, H. Inoue, and T. Dotani}
               {Origin of the ``Disk-Line'' Spectrum Feature in NGC~4151}
\Received{2001/05/22}
\Accepted{2002/03/15}

\title{Origin of the ``Disk-Line'' Feature in the X-Ray Energy Spectrum
        of a Seyfert Galaxy, NGC~4151}
\author{Kazuki \textsc{Takahashi},
         Hajime \textsc{Inoue},
         and
         Tadayasu \textsc{Dotani}}
\affil{The Institute of Space and Astronautical Science\\
        3-1-1 Yoshinodai, Sagamihara, Kanagawa 229-8510}
\email{kazuki@astro.isas.ac.jp, inoue@astro.isas.ac.jp, 
       dotani@astro.isas.ac.jp}

\KeyWords{galaxies: individual (NGC~4151) --- galaxies: Seyfert 
          --- X-rays: galaxies --- X-rays: individual (NGC~4151) 
          --- X-rays: sources}

\maketitle

\begin{abstract}
We have studied the origin of the broad and skewed feature at 4.5--7.5~keV
in the energy spectra of NGC~4151 using the ASCA and RXTE data.
The feature consists of a narrow peak at 6.4~keV and a broad wing 
extended between 4.5--7.5~keV\@.
An analysis of the long-term variations revealed that the feature 
became variable only on a time scale longer than $1.5 \times 10^6$~s.
Through a comparison with the continuum variabilities, we 
found that the emission region of the excess flux at 4.5--7.5~keV
has an extent of $10^{17}$~cm.
The broad and skewed feature at 4.5--7.5~keV may be
explained by the so-called ``disk-line'' model.
If so, the size of the line-emitting region, $10^{17}$~cm, should be
equal to several or ten-times the Schwarzschild radius of the
central black hole.
This results in a black hole mass of $10^{11}$~$M_\odot$,
which may be too large for NGC~4151.
We propose an alternative explanation for the broad and skewed feature,
i.e.\ a ``reflection'' model, which can also reproduce the overall
energy spectra very well.
In this model, cold matter with a sufficiently large column density is
irradiated by X-rays to produce a reflected continuum, which constitutes
the broad wing of the feature, and narrow fluorescent lines.
The equivalent width of the iron fluorescent line ($\sim$2~keV)
and the upper limit of its width ($\sigma < 92$~eV) are also
consistent with this model.
 From these results and considerations, we conclude that the ``disk-line''
model has difficulty to explain the spectral variations of NGC~4151, and the
reflection model is more plausible.

\end{abstract}

\section{Introduction}
Active Galactic Nuclei (AGN) are the most luminous objects in the universe.
They emit large amounts of radiation over a wide range of wave-bands
(from radio to $\gamma$-rays), and sometimes produce relativistic jets.
It is a goal of modern astronomy to understand the origin
of the extreme activities in AGN\@.

One of the key observational clues in the X-ray band to understand 
AGN is the iron line profile in the energy spectrum \citep{fab00}.
A broad and skewed line feature around 5--7~keV was clearly
detected by ASCA for the first time in an energy spectrum of
the Seyfert galaxy MCG--6-30-15 \citep{tan95}.
Since then, a similar feature was detected by ASCA
from several Seyfert galaxies (e.g.\ \cite{nan97}).
The broad and skewed feature has been interpreted as an iron fluorescent
line originating from the innermost region of an accretion disk
(\cite{fab95}, so-called ``disk-line'' model).
In this model, a broad and skewed feature is explained by the
combination of Doppler broadening due to the relativistic motion
of the line emitting matter and gravitational redshift due to
the central massive black hole.

If the broad and skewed feature in the energy spectra is really
a ``disk-line'', we can determine the inclination angle of the 
accretion disk through an analysis of the line profile \citep{fab89}.
However, the thus-obtained inclination angle is sometimes inconsistent
with that estimated with other methods (e.g.\ \cite{nis98}).
It may also be inconsistent with an expectation from the unified scheme,
in which the accretion disk in Seyfert 2 galaxies should have
an edge-on geometry.
ASCA observations have revealed the presence of a broad and
skewed line feature in several Seyfert 2 galaxies.
The inclination angles have been deduced for these Seyfert galaxies from 
an analysis of the ``disk-line'', and always found to be around
$30^\circ$ \citep{tur98}.
This inclination angle is unexpectedly small.
This problem may be partially solved if we assume that the line is
a composite and consists of a narrow line centered at 6.4~keV
and a disk line with intermediate inclination angle \citep{wea98}.
However, we still need an observational confirmation on such
a composite model.

NGC~4151 is a bright, nearby Seyfert galaxy (type 1.5; \cite{ost76};
for a review see \cite{ulr00}).
The presence of the broad and skewed line feature in 4.5--7.5~keV
has been known through ASCA observations \citep{yaq95,wan99}.
Thus, NGC~4151 is one of the best targets to precisely study the line feature.
A ``disk-line'' analysis of NGC~4151 has resulted in a face-on geometry
of the accretion disk \citep{yaq95}, while observations of the
[O III] $\lambda$5007 image with the Hubble Space Telescope indicate
an inclination angle of $60\pm5^\circ$ \citep{eva93}.
This inconsistency might be partially resolved if we assume two disk
lines and an additional narrow peak centered at 6.4~keV to reproduce
the profile \citep{wan99}.
However, it is not clear how the two disk lines with different inclination
angles can be produced simultaneously.

In this paper, we analyze the broad and skewed feature observed from
NGC~4151 using a model-independent method as much as possible, 
and try to identify the origin of the feature.
For this purpose, we analyze the long observations of NGC~4151 made with
ASCA (in 2000) and RXTE (in 1999), while focusing on the spectral 
variations on various time scales.
The results from another set of long ASCA observations of NGC~4151 have
been presented by \citet{wan01}.

\section{Observations}

\subsection{ASCA}

The primary ASCA data set used for the present analysis was
acquired from 2000 May 13 through 25.
Concerning the instrumentation on board ASCA, see the following 
references: \citet{tan94} for a general description of ASCA,
\citet{ser95} for the X-ray telescope, \citet{oha96} and \citet{mak96}
for the Gas Imaging Spectrometer (GIS), and \citet{bur94} for the
Solid-state Imaging Spectrometer (SIS)\@.
In the observations, SIS was operated in the 1-CCD Faint
mode and GIS in the standard PH mode.
A lower level discriminator was applied to SIS at 0.48~keV to avoid 
telemetry saturation due to flickering pixels.
The data selection criteria for SIS and GIS are summarized in
table~\ref{tbl:crit}.
When the elevation angle, i.e.\ the angle between
the field of view and the Earth limb, was less than $5^\circ$,
the data were discarded because they were affected by scattered X-rays
from the day Earth or by the absorption due to the atmosphere.
Some of the optical light leakage is known in SIS, and the SIS data
obtained when the elevation from the day Earth was less than
$25^\circ$ were discarded.
The data with telemetry saturation are also discarded to avoid
any deterioration of the detection efficiency.
The net exposure of SIS was about 350~ks, which spans 
$1.1\times10^6$~s, and an average count rate of SIS was
0.8~count~s$^{-1}$ after data screening and reduction.
An X-ray source was detected with both SIS and GIS at 
a position consistent with the optical position of NGC~4151 
within the positional accuracy of ASCA\@.
A BL Lac object (1207+39W4; \cite{arp97}) was detected in the field of 
view of ASCA about \timeform{4.7'} north of NGC~4151.
Contamination by this source was found to be negligible above 2~keV,
which was used for the current data analysis.
According to the Chandra observations, a faint source was also present
at \timeform{2.2'} from NGC~4151 to the southwest \citep{yan01}.
This source was not resolved with ASCA\@.
Contamination by this source was completely negligible.

The energy spectra and light curves of each detector were 
accumulated from a circular region centered on the source.   
The extraction radius of SIS was set to \timeform{3.6'}, 
which is the maximum radius fit in both chips, on the whole.
The background spectra of SIS are accumulated from the source
free region on the chip.
The extraction radius of GIS was set to 6~arcmin, because various 
calibrations of GIS had been done with this standard extraction radius.
The background data of GIS were accumulated from a circular 
region opposite to the target position against the bore-sight 
axis, excluding the source region.

We usually obtained 4 sets of energy spectra from the ASCA data, two 
from SIS and the other two from GIS, for a single set of observations.
To make the analysis simple, we summed up the two energy spectra from SIS
and the two corresponding response matrices, respectively, which produced
a single SIS spectrum with the corresponding response.
Similarly, we summed up the two GIS spectra and the corresponding responses,
respectively.    The summed SIS and GIS spectra were fitted with a model
spectrum simultaneously.    
The simultaneous fitting made a full utilization of the ASCA data 
in the sense that the higher detection efficiency below 5~keV
and better energy resolution of SIS, and the higher detection efficiency
above 5~keV of GIS are all reflected in the best-fit spectral parameters.
Because there is 3\% uncertainty of flux normalization between SIS
and GIS, a systematic error of 3\% was conservatively added to the 
SIS energy spectra in the simultaneous fitting.
Although a decrement of SIS detection efficiency below 2~keV was 
reported by the instrument team (T. Yaqoob, private communications), 
we did not include the effect in the analysis because the spectral 
fittings were carried out only for the energy range above 2~keV\@.

\subsection{RXTE}

Data obtained by the Proportional Counter Array (PCA) on board the Rossi 
X-ray Timing Experiment (RXTE) were also used in the present analysis.
Details of RXTE and PCA can be found in \citet{jah96}.
RXTE observations of NGC~4151 were carried out every $\sim$5~days
from the beginning of 1999 to study the long-term variations of the
X-ray flux.
We used the data obtained from 1999 January 1 through July 24,
which were publicly available from the archive in 
HEASARC\footnote{\tt http://heasarc.gsfc.nasa.gov/} at the time of
data analysis.
Reduction of the PCA data was carried out using the standard 
procedure \citep{ede99}.
In the analysis, we discarded data obtained when
the elevation from the Earth limb was less than $10^\circ$,
data obtained within 30~minute after the satellite's passage of SAA,
data obtained when the pointing offset was greater than \timeform{0.02D},
and data when the count rate of electrons was greater than 0.1.
The background spectra were calculated using a faint-source model 
according to a method described in the RXTE Cook Book.\footnote{\tt 
http://heasarc.gsfc.nasa.gov/docs/xte/recipes/cook\_book.html}

\section{Analyses and Results}
\subsection{ASCA Data}
\begin{figure}
\begin{center}
\FigureFile(85mm,60mm){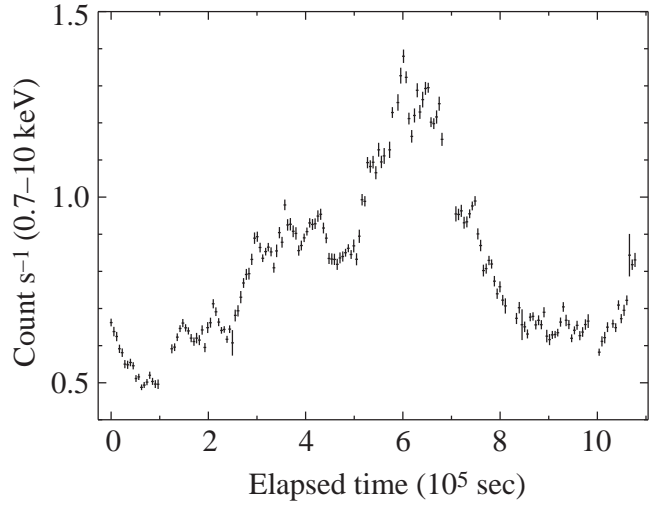}
\end{center}
\caption
{X-ray light curve of NGC~4151 in 5.6~ks time bin calculated with 
the ASCA GIS data obtained during 2000 May 13--25.
The X-ray flux varied by a factor of 2--3 during the observations. }
\label{fig:ascaLC}
\end{figure}

Figure~\ref{fig:ascaLC} shows the X-ray light curve at 0.7--10~keV
from NGC~4151 obtained with ASCA GIS on 2000 May 13--25.
Here, the data are shown in 5.6~ks binning.
This 5.6~ks bin-time corresponds to the orbital period of ASCA
around the Earth.
We chose this bin time because: (1) time variations shorter than the
orbital period were not significant, (2) data gaps are usually much
shorter than the orbital period and we can get continuous light curve,
(3) possible background variations, which are correlated with the
orbital period, could be largely reduced.
Note that data gaps were produced mainly by the earth occultation of
the source, high background regions on the earth, and down-link to the
ground stations.
X-ray flux variations of a factor of 2--3 are clearly seen in the
figure.

\begin{figure}
\begin{center}
\FigureFile(85mm,60mm){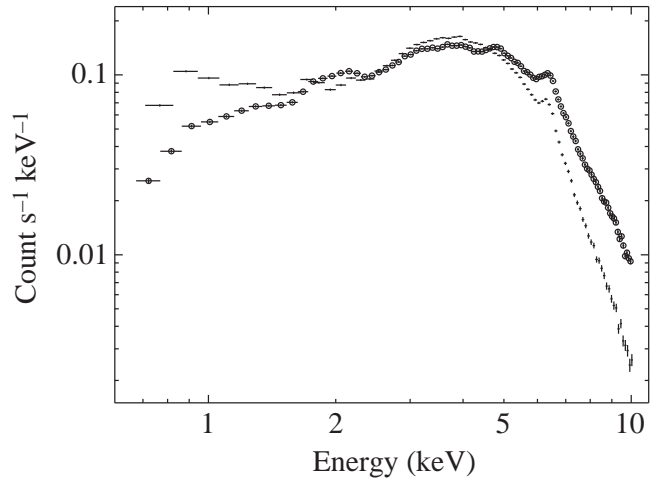}
\end{center}
\caption{Average energy spectra (folded with the detector response) of
NGC~4151 obtained with ASCA in 2000 May.   The spectra of GIS
(crosses with open circles) and SIS (bare crosses) are shown separately.
A prominent line like structure can be seen around 6.4~keV in both energy
spectra. }\label{fig:aveSpe}
\end{figure}

\begin{figure}
\begin{center}
\FigureFile(150mm,120mm){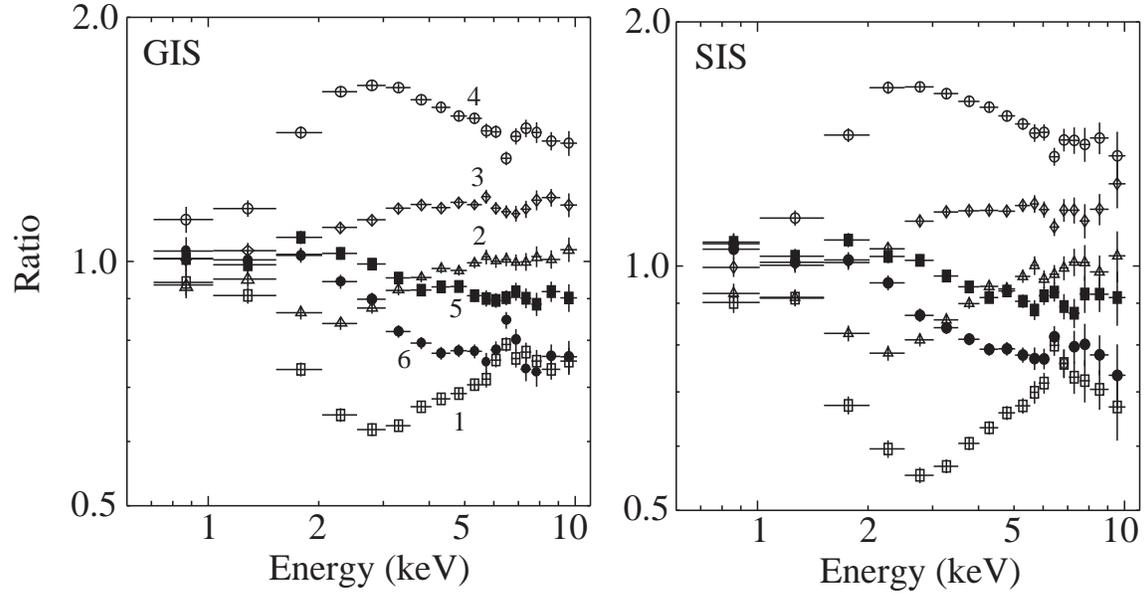}
\end{center}
\caption{Ratios of the time-resolved energy spectra (obtained every
$1.8\times10^5$~s) to the average energy spectrum.
The left panel shows those obtained from GIS and the right panel 
from SIS\@.    The ratios are consistent between GIS and SIS\@.
Time sequence of the spectra is indicated in the left panel.
}
\label{fig:pharatio}
\end{figure}

We next studied the energy dependence of the flux variations.
For this purpose, we calculated the ratios of the time-resolved energy
spectra to the average energy spectrum.
Considering the statistics of the data and the amplitude of the 
time variations, the time-resolved energy spectra were calculated 
every $1.8\times10^5$~s and six spectra were obtained in total.
We also calculated the average spectra for GIS and SIS, each of 
which correspond to the average of 6 time-resolved spectra, 
as shown in figure~\ref{fig:aveSpe}.
The ratios of the six time-resolved spectra to the average spectrum are
presented in figure~\ref{fig:pharatio}.
Large time variations in both the flux and the shape can be seen at 2--5~keV,
while the ratios are almost flat at 0.7--1.5~keV and 7--10~keV\@.
The flat spectral ratios mean that the shape of the energy
spectra does not vary at 0.7--1.5~keV and 7--10~keV\@.
Furthermore, those at 0.7--1.5~keV are always close to unity.
This indicates that the energy spectra change neither their shapes nor
the fluxes in this energy range.
It should be noted that a local structure is seen just around 6.4~keV and
the ratios are closer to unity only in this narrow energy band.
At least three spectral components may be required to reproduce
these time variations of the spectral ratios: a stable soft component
below $\sim$2~keV, a variable hard component above $\sim$2~keV,
and a relatively stable component around 6--7~keV\@.
Because we are mostly interested in the feature around 6--7~keV,
we do not discuss the soft stable component hereafter by limiting the
energy range in the spectral analyses to above $\sim$2~keV\@.

\begin{figure}
\begin{center}
\FigureFile(85mm,60mm){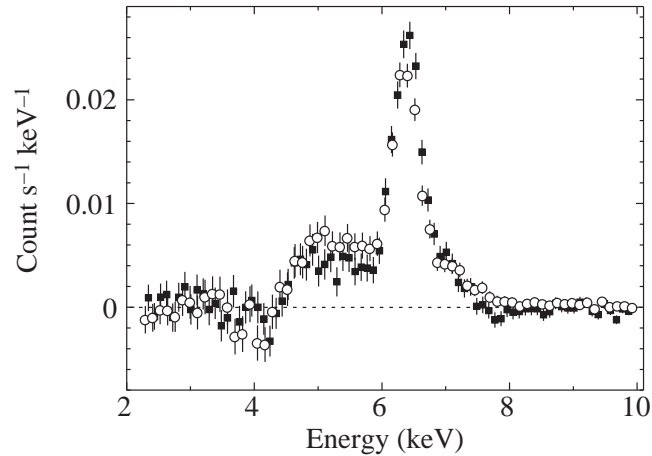}
\end{center}
\caption{Excess feature in 4--8~keV after subtracting the continuum
model.   The continuum model was determined by fitting the
average spectrum in two energy ranges, 2.2--4.0~keV and 8--10~keV\@.
The data with open circles and those with filled squares show
the results for SIS and GIS, respectively. }\label{fig:fe}
\end{figure}

In order to study the nature of the relatively stable component around
6--7~keV, we need to separate the component from the underlying
variable hard component.
We first determined the shape of the variable hard component by performing
a model fit to the average spectrum at 2--4~keV and 8--10~keV after
masking the 4--8~keV range.
The results show that a simple model of a power law with a single
absorption cannot reproduce the spectrum, and that a ``dual absorber''
model is necessary.     The need for a dual absorber for the
ASCA spectrum was already pointed out by \citet{wea94}.
The ``dual absorber'' model is expressed by the following formula:
\begin{equation}
F_{\rm dual}(E) = A \left\{ C_{\rm F} \exp[-\sigma(E) N^{1}_{\rm H}]
        + (1 - C_{\rm F}) \exp[-\sigma(E) N^{2}_{\rm H}] \right\} E^{-\Gamma},
\label{eq:dualAbs}
\end{equation}
where $A$ is a normalization factor, $\Gamma$ the photon index, 
$N_{\rm H}$ the column density, $\sigma(E)$ the photo-electric 
absorption cross-section, and $C_{\rm F}$ the source covering fraction 
($0<C_{\rm F}<1$).
The superscript to $N_{\rm H}$ indicates two different column densities.
In the following analysis, the photo-electric absorption cross-sections 
compiled by \citet{mor83} are considered.
The thus-determined continuum spectrum was subtracted from the 
average spectrum; the residual feature at 4--8~keV is shown
in figure~\ref{fig:fe}.
The presence of a narrow peak at 6.4~keV and a broad feature skewed 
toward the lower energy can be clearly seen.

\begin{figure}
\begin{center}
\FigureFile(150mm,130mm){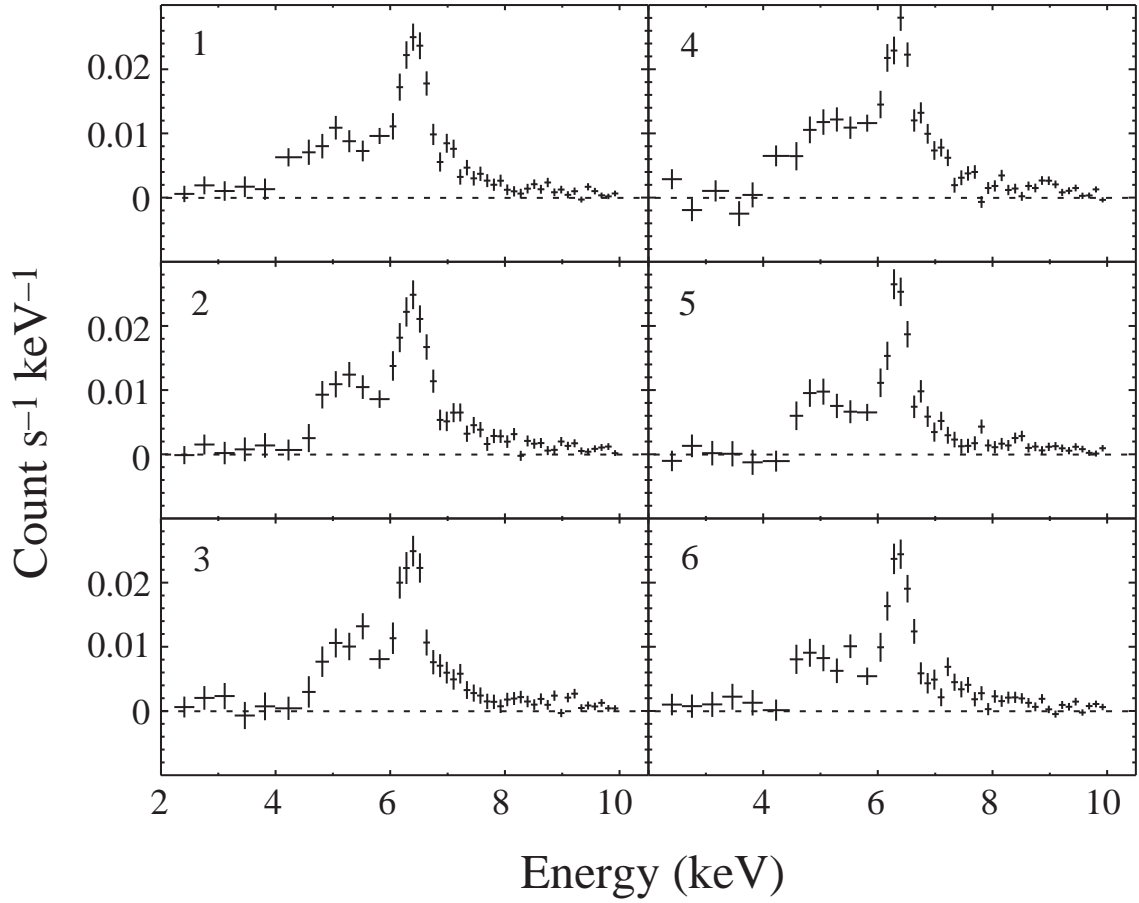}
\end{center}
\caption{Profiles of the excess features in 4--8~keV of the six
time-resolved SIS spectra after subtracting the continuum model. The time
sequence is indicated in the panels.
}\label{fig:feIndiv}
\end{figure}

We then studied the time variations of the residual feature 
at around 6.4~keV\@.
For this purpose, we used the 6 sets of SIS spectra, which have a 
better energy resolution than the GIS spectra.
The continuum shape was determined separately for the 6 sets of spectra
using the same method for the average spectrum.
We set $N_{\rm H}^{2}$ in equation~(\ref{eq:dualAbs}) to 0 to simplify
the model fitting, considering the statistics of the data.
We confirmed that setting $N_{\rm H}^{2}$ to zero did not
change the continuum shape at 4--8~keV more than 3\% while taking 
the 4th data set (those with the largest flux) as an example.
The model was found to be acceptable for all 6 sets of the spectra 
with the largest value of $\chi^2_\nu/{\rm d.o.f.} = 0.97/61$.
When the continuum model was interpolated to the 4--8~keV band, 
an excess feature could be clearly seen in all 6 sets of the 
energy spectra.
We show them in figure~\ref{fig:feIndiv} after subtracting the continuum
model from the observed spectra.   
A prominent narrow peak at around 6.4~keV,
and broad features at 4.5--6.0~keV (hereafter referred to as a red wing) 
and at 6.8--7.5~keV (a blue wing) are clearly seen in all 6 data sets.

\begin{figure}
\begin{center}
\FigureFile(150mm,100mm){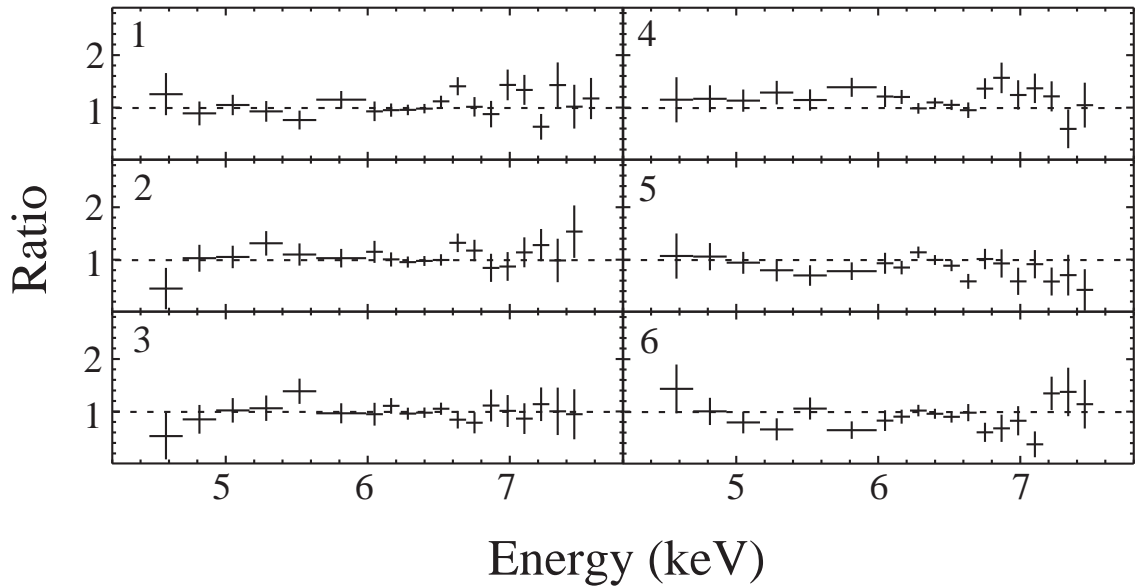}
\end{center}
\caption{Ratios of the excess features (see figure~\ref{fig:feIndiv})
to their average deduced from the ASCA SIS data.
All of the ratios are consistent to be constant.
This means that the excess feature in 4--8~keV show little time variations.
}\label{fig:ExcessPHA}
\end{figure}

In order to study any profile change of the broad and skewed feature,
we made the ratios of each of the 6 excess features to their average.
The ratios were calculated by dividing the individual excess feature 
(after subtracting the interpolated continuum) by the excess feature
in the average spectrum (figure~\ref{fig:fe}).
The results are shown in figure~\ref{fig:ExcessPHA}.
By performing a $\chi^2$ test for the hypothesis that the ratios have no
energy dependence, we found that the 6 sets of ratios are consistent
with being constant in terms of the energy, and that the time variation
of the profile of the broad and skewed feature is not significant.
At the same time, we calculated the flux ratio of the red wing 
(4.5--6.0~keV) to the narrow core (6.0--6.8~keV) and that of the blue 
wing (6.8--7.5~keV) to the narrow core for the 6 sets of SIS spectra.
No significant time variation was found at the 90\% confidence limit 
in the two sets of ratios ($\chi^{2}_{\nu}/{\rm d.o.f.} = 1.22/5$ 
at most).
The standard deviation to the mean for the flux ratio of the red
wing to the narrow core was calculated to be 20\% at most.
Although the time variation is not significantly found, a variation
of less than 20\% in amplitude is not rejected.

\subsection{RXTE Data}

\begin{figure}
\begin{center}
\FigureFile(85mm,60mm){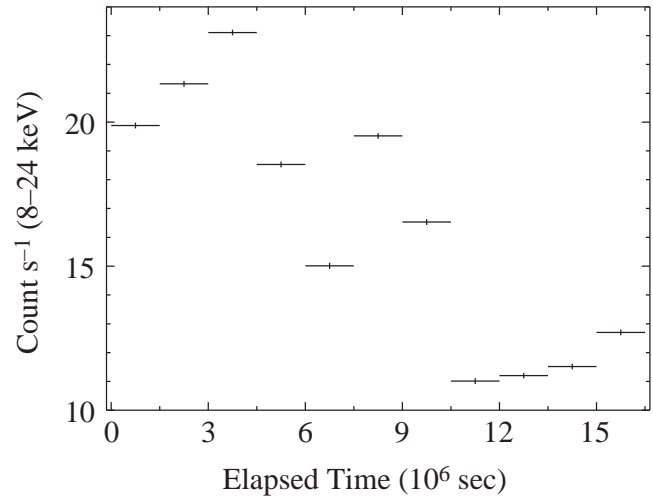}
\end{center}
\caption{Light curve of NGC~4151 in 8--24 keV calculated from the
RXTE PCA data obtained from 1999 January 1 through July 24.
Significant flux variations are clearly seen.
}\label{fig:rxteLC}
\end{figure}

\begin{figure}
\begin{center}
\FigureFile(85mm,60mm){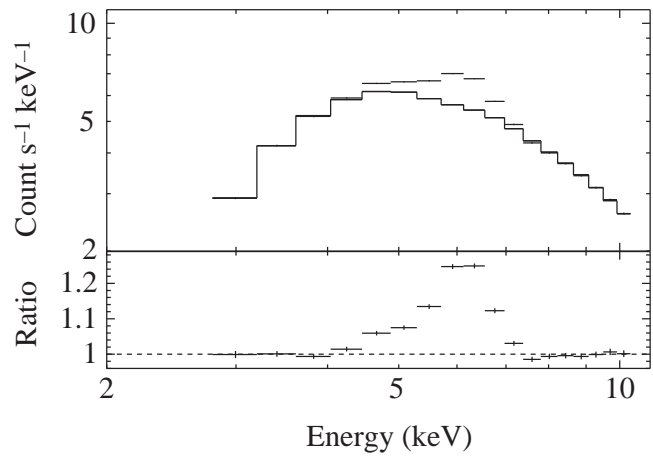}
\end{center}
\caption{PCA spectrum of NGC~4151 averaged over the RXTE observations from
1999 January 1 through July 24, (crosses in the upper panel) together
with the best fit continuum model to the spectrum after masking 4--8~keV
range (histogram in the upper panel).
The lower panel shows the ratio of the observed spectrum to the continuum
model. }\label{fig:rxteAve}
\end{figure}

We analyzed the RXTE PCA data in order to study the spectral variations
on a time scale longer than the $1.1\times10^6$~s covered by ASCA\@.
We show in figure~\ref{fig:rxteLC} the light curve calculated from
the PCA data obtained from 1999 January 1 through July 24.
Significant flux variations can be clearly seen.
The PCA spectrum averaged over the RXTE observations is shown in
figure~\ref{fig:rxteAve}.
As was done for the ASCA spectra, we fit a power law modified by
``dual absorption'' ($N_{\rm H}^{2}=0$) to the average spectrum 
in 2.8--10~keV after masking the 4--8~keV energy band.
We also checked how it affected the continuum shape to set $N_{\rm H}^{2}$
to zero, and found that there was no noticeable impact if the fit 
range was restricted to 2.8--4 and 8--10 keV\@.
The ratio of the observed spectrum to the
continuum model is shown in the lower panel of figure~\ref{fig:rxteAve}.
We can clearly see the presence of a broad and skewed feature at
4--8~keV again.

\begin{figure}
\begin{center}
\FigureFile(150mm,120mm){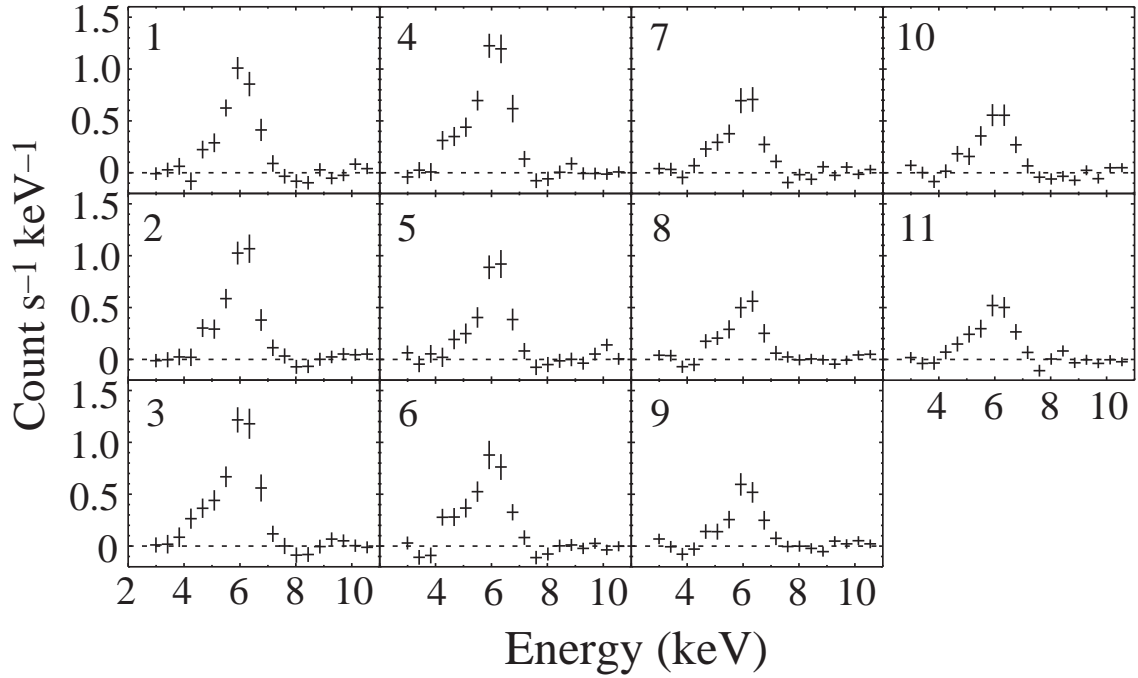}
\end{center}
\caption{Residuals of the 11 PCA spectra after subtracting the continuum
models.  The residuals were calculated from spectra accumulated
for $1.5\times10^6$~s each by subtracting the continuum model determined
by excluding the 4--8~keV band.   A significant excess flux over the 
continuum model is always seen in the 4.5--7.5~keV band.
}\label{fig:rxteIndiv}
\end{figure}

\begin{figure}
\begin{center}
\FigureFile(150mm,120mm){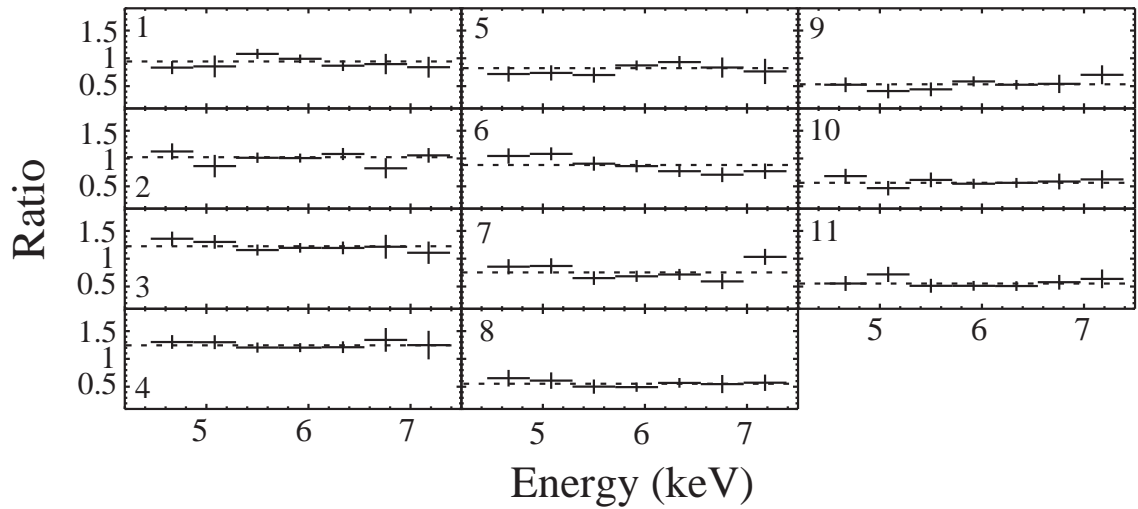}
\end{center}
\caption{Ratios of the 11 residual spectra (see figure~\ref{fig:rxteIndiv})
to their average.   The ratios are all consistent with being constant,
although the average flux level (indicated by the broken line) changes
among the 11 ratios.
}\label{fig:rxteRatio}
\end{figure}

Next, in order to search for the time variations of the broad and skewed
feature on time scales longer than $1.1\times10^6$~s, we divided the
data into 11 subsets on a time bin of $1.5\times10^6$~s and 
calculated an energy spectrum for each subset of data.
We then fit a power law modified by ``dual absorption'' to the 11 
spectra in 2.8--10~keV while excluding the 4--8~keV band again.
The fit was acceptable for all of the spectra.   The residuals of 
the 11 spectra after subtracting the best-fit continuum model
are shown in figure~\ref{fig:rxteIndiv}.
A significant excess flux over the continuum model is always seen
at 4.5--7.5~keV\@.   The excess has a peak at around 6.0--6.5~keV 
as well as a significant tail feature on the lower energy side.
Then, we again investigated the time variations of the spectral shape of
the excess by comparing the 11 excess spectra with their average.
The spectral ratios of the 11 spectra to their average are shown
in figure~\ref{fig:rxteRatio}.
We performed $\chi^2$ tests to check whether or not each of the 11
ratio-spectra is consistent with being flat.
A flat model was found to be acceptable for all 11 ratios
with $\chi^{2}_{\nu}/{\rm d.o.f.}<1.23/7$, although the flux level 
of the excess shows a significant variation.
This indicates that, although the excess flux at 4.5--7.5~keV 
is variable on a time scale longer than $1.5\times10^6$~s,
the profile does not change significantly.

\section{Plausible Models}

In the previous section we discussed the time variability of the flux
and the profile of the broad and skewed feature consisting of a narrow
peak at 6.4~keV and red/blue wings in a way fairly independent of
any spectral models.
In this section, we introduce two different model functions to
reproduce the profile: one is the so-called ``disk-line'' model,
and the other is a ``reflection'' model.

\subsection{Disk-Line Model}

It is widely accepted that the broad and skewed feature at around 
5--7~keV often seen from AGN is composed of gravitationally 
redshifted iron-fluorescent lines from matter orbiting in an 
accretion disk very close to the central black hole~\citep{tan95, fab95}.
Hence, we first tried to fit the broad line-like feature observed
from NGC~4151 with the disk-line model.
We adopted a spectrum described by equation~(\ref{eq:dualAbs}) as
the continuum, where $N_{\rm H}^2$ was set to 0 to simplify the model.
The outer and inner radii of the accretion disk were set to
1000~$r_{\rm Sch}$ and 10~$r_{\rm Sch}$, respectively, where
$r_{\rm Sch}$ is the Schwarzschild radius.
The emissivity of the line emission is assumed to have a power-law
dependence on the radius with the index of $-$2 \citep{fab89}.

The model fitting was carried out in the energy range of 2.2--10~keV
simultaneously to the SIS and GIS data for each of the 6 sets of
the ASCA energy spectra.
The photon index was fixed to 1.55, which was the average value
when we fit a power law with a partial covering absorber to the
spectra after masking 4--8~keV\@.
The model was found to be acceptable for all of the spectra at 
2.2--10.0~keV with $\chi^2_\nu/{\rm d.o.f.}=(0.99\!-\!1.10)/140$.
The best-fit parameters are listed in table~\ref{tbl:ascaPar}.
The inclination angles and the line center energies obtained from the
spectral fits to the 6 spectra are not significantly different from one
another.  The reduced $\chi^2$ values for the hypothesis that the 6 values
are constant were 1.35 and 0.89 for 5 d.o.f.\ for the inclination
angle and the line center energy, respectively.
This means that these two parameters are consistent with being constant.
The average inclination angle and the line center energy are
$20^\circ$ and $6.44$~keV, respectively.

We applied the same model to the 11 energy spectra of PCA at 2.8--24~keV\@.
However, the fit was not acceptable with $\chi^2_\nu/{\rm d.o.f.} > 1.89/46$.
The deviation of the model from the observed spectra became noticeable
above 16~keV\@.  Hence, we replaced the power law in
equation~(\ref{eq:dualAbs}) with a broken power law as described below:
\begin{equation}
F_{\rm bkn}(E) = \left \{
                      \begin{array}{ll}
                              A \, E^{-\Gamma_1} & (E \le E_{\rm bk})    \\
                              A \, E_{\rm bk}^{\Gamma_2 - \Gamma_1} \,
                                   E^{-\Gamma_2} & (E \ge E_{\rm bk}), \\
                      \end{array}
                        \right.    \label{eq:bknPow}
\end{equation}
where $A$ is a normalization, $\Gamma_1$ and $\Gamma_2$ are photon indices,
and $E_{\rm bk}$ is a break energy at which the spectral index changes.
Note that the broken power-law is required due to the apparent change
in the spectral slope above $\sim$16~keV, not because we set
$N_{\rm H}^2$ to zero.

The 11 energy spectra of PCA were all reproduced by the new model with
$\chi^2_\nu/{\rm d.o.f.}=(0.99\!-\!1.48)/46$.
The best-fit parameters are listed in table~\ref{tbl:rxtePar}.
We checked the time variation of the 11 values
of each parameter by performing a $\chi^2$ test.
The photon indices, $\Gamma_1$ and $\Gamma_2$, and the break energy,
$E_{\rm bk}$, are found to have not changed significantly during
the observational period.
We cross-checked the spectral slope above the break energy, which
was found to be $\Gamma_2 = 1.88\pm0.17$, using the data from another
detector HEXTE (High Energy X-ray Timing Experiment) on board RXTE\@.
HEXTE consists of two clusters of 4 Na\hspace{0.5ex}I/Cs\hspace{0.5ex}I 
phoswich scintillation detectors sensitive to X-rays from 15 to 250~keV\@.
The photon indices obtained from a simultaneous fit of a power-law model
to the 11 sets of the PCA spectrum (18--24~keV) and HEXTE spectrum
(18--100~keV) are found to fall around $1.8\pm0.2$.
This is consistent with $\Gamma_2 \sim 1.9$.
It is noteworthy that, although the disk line model can always reproduce 
the excess feature around 5--7~keV, the continuum spectrum needs a 
break at around 17~keV to reproduce the energy spectrum up to 100~keV\@.

\subsection{Reflection Model}

In the disk-line model, the peak at 6.4~keV is considered to be a blue part
of fluorescent lines coming from matter orbiting in an accretion disk
very close to the central black hole.   However, the peak energy is just the
K$_{\alpha}$ line of a neutral, or a lowly ionized iron, in the rest frame;
furthermore, the profile around the peak can be reproduced by a single
narrow line.   Hence, it would be more natural to consider that the peak at
6.4~keV is a single fluorescent iron line coming from a region free from
relativistic effects.   If the line is emitted through the fluorescent
process as a result of X-ray illumination on relatively cold matter far
outside the central X-ray source, X-ray reflection at the surface of the
matter should take place simultaneously with the fluorescent line emission.
The X-rays reflected by the cold matter emitting the fluorescent lines
could be a possible origin of the broad red and blue wings accompanied
by the narrow line feature on both sides.
Because X-ray reflection is due to Thomson scattering, the reflected
X-rays should have the same spectrum as the illuminating X-rays in a typical
X-ray band, but suffer from absorption when they pass through the cold mater.
Thus, the reflected component may be approximated by applying cold-matter
absorption to the continuum model.

In order to test this possibility, we introduced the following 
model spectrum:
\begin{equation}
F_{\rm ref}(E)  =  P(E,\Gamma) \{C_{\rm F}^1 \exp[-\sigma(E)
                  N_{\rm H}^1] + 1-C_{\rm F}^1\} \exp[-\sigma(E)
                  N_{\rm H}^2] 
                  + I(E_{\rm line})  + C_{\rm F}^2 P(E,\Gamma)
                   \exp[-\sigma(E) N_{\rm H}^3].   \label{eq:ref}
\end{equation}
Here, the first term is the same as equation~(\ref{eq:dualAbs}),
representing continuum X-rays directly coming from the central X-ray
source.   The second term represents the fluorescent iron line at 6.4~keV
and $I(E_{\rm{line}})$ is a Gaussian function.
The line width is assumed to be zero.
The third term is for the reflection component.
The efficiency of the X-ray reflection is represented by $C_{\rm F}^2$,
which corresponds to a covering fraction of the X-ray reflector
optically thick for Thomson scattering.
The function, $P(E,\Gamma)$, is a power law with a photon index of $\Gamma$.
The photon index is optimized in the course of the fitting, because
it may be affected by the introduction of a highly absorbed component.
The column densities and covering fractions are also optimized.

\begin{figure}
\begin{center}
\FigureFile(85mm,60mm){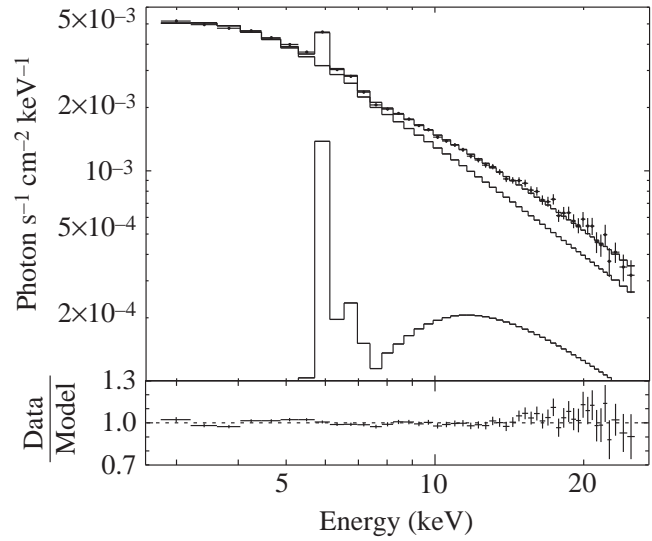}
\end{center}
\caption{Example of the fits of the ``reflection'' model to
the PCA spectra.  The crosses represent the observed spectrum, while
the histograms represent the two model-components, the direct 
power-law component 
and the reflected power law $+$ the narrow line component.
}\label{fig:rxteRef}
\end{figure}

Model fitting was firstly carried out to the 11 PCA spectra in the energy
range of 2.8--24~keV\@.
In this energy range, because the effect of absorption by $N_{\rm H}^2$ 
was negligible, we omitted this term.
The fit was acceptable for all energy spectra with
$\chi^2_\nu/{\rm d.o.f.}=(0.87\!-\!0.99)/46$.
An example of the results of the model fitting is shown in
figure~\ref{fig:rxteRef} and the best-fit parameters are listed in
table~\ref{tbl:rxteRef}.
The results show that 40--50\% of the direct flux is reflected.
We checked the constancy of the 11 values of each spectral 
parameter with a $\chi^2$ test.
We found that the photon index and the absorption column of 
the reflected component, $N_{\rm H}^3$, are both consistent with
being constant ($\chi^2_\nu/{\rm d.o.f.} = 0.98/10$, and 1.25/10, 
respectively).
We confirmed that the profile of the broad and skewed feature did not
change significantly during the RXTE observations.

\begin{figure}
\begin{center}
\FigureFile(85mm,60mm){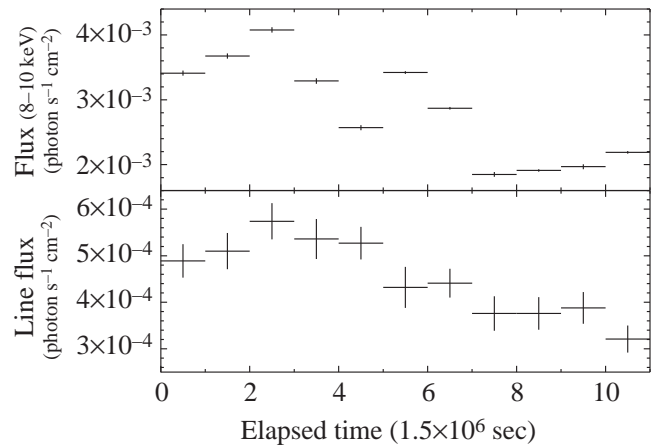}
\end{center}
\caption{Light curves of the continuum flux (upper panel) and the iron line
flux (lower panel) on a $1.5\times 10^6$~s time bin. These light curves
were obtained form RXTE observations.   The continuum flux was calculated
in 8.0--10 keV\@.}\label{fig:rxteFe}
\end{figure}

The weighted mean of the 11 photon indices is $\Gamma = 1.92\pm0.06$.
This is comparable to $\Gamma =1.8\pm0.2$, which we obtained from a
simultaneous fitting to the 11 sets of the PCA energy spectra at
18--24~keV and the HEXTE energy spectra at 18--100~keV\@.
If we extrapolate the best-fit model at 2.8--24~keV to 100~keV,
no significant discrepancy is recognized between the model and the
observed energy spectra with HEXTE\@.
This indicates that the intrinsic spectrum emitted from the central source
should be a single power law with an index of $\sim$1.9, and should extend
up to 100~keV without a break, if the ``reflection'' model explains
the broad and skewed feature at around 5--7~keV\@.
Although the best-fit line energies are found to be consistent with being
constant, the mean line center energy is $E_{\rm line}=6.1\pm0.1$~keV\@.
This is slightly lower than the center energy of the iron fluorescent line
from neutral matter (6.4~keV)\@.
However, it is known that there is a systematic uncertainty of about
2--3\% in the energy scale of the RXTE instrument.
Thus, the line energy can still be interpreted as the iron fluorescent line.
The time variation of the line flux is shown together with that of the
continuum flux at 8--10~keV in figure~\ref{fig:rxteFe}.
The time variation of the line flux is clearly seen in a time scale
of $10^6$--$10^7$~s.

\begin{figure}
\begin{center}
\FigureFile(85mm,60mm){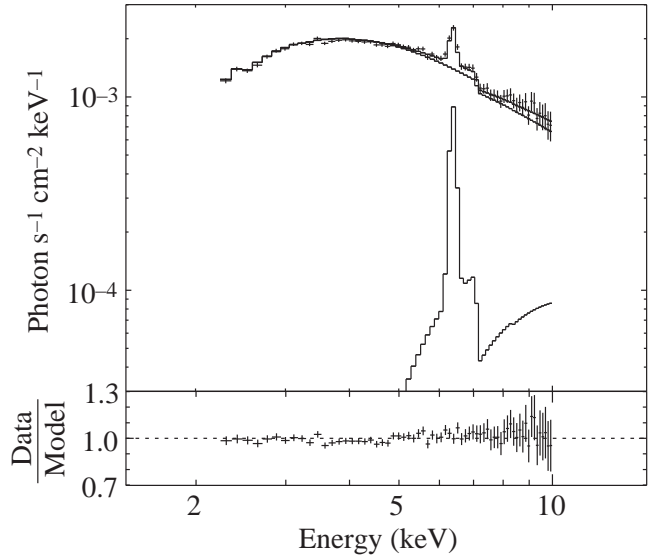}
\end{center}
\caption{Spectral fit of the ``reflection'' model to the ASCA SIS 
spectrum of NGC~4151.   The crosses represent the observed spectrum.
The histograms are two model-components: the direct power-law
component and the reflected power law $+$ the narrow line component.
The lower panel shows the ratio of the observed spectrum to the model-spectrum.
} \label{fig:ascaRef}
\end{figure}

\begin{figure}
\begin{center}
\FigureFile(85mm,60mm){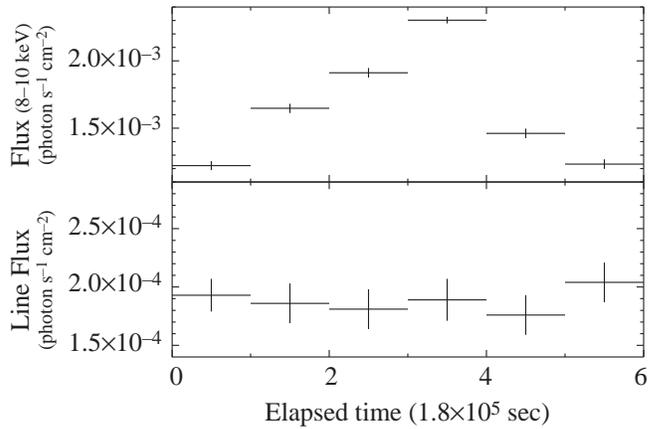}
\end{center}
\caption{Light curves of the line flux (lower panel) and the continuum flux
(upper panel) with 1~$\sigma $ error are plotted in $1.8\times10^5$~s
binning.   These curves are obtained from the ASCA data.  The continuum
flux was calculated at 8.0--10~keV\@.   The line flux does not show
significant time variations.
} \label{fig:ascaLCfe}
\end{figure}

Next, we also applied the model function (equation~\ref{eq:ref}) 
to the ASCA data.   
Because the energy range of the ASCA detectors is limited below 
10~keV, some of the model parameters were not well constrained.
Hence, we fixed the following parameters in the fitting: $\Gamma$ ($=1.9$),
$N_{\rm H}^3$ ($=10^{24}$ atom~cm$^{-2}$) and $C_{\rm F}^2$ ($=47$~\%).
Other parameters were optimized during the course of fitting.
The model fitting was carried out to each of the 6 sets of the ASCA
data, simultaneously to the SIS and GIS spectra.
Fittings of this model were all acceptable with
$\chi^{2}/{\rm d.o.f.}=(0.94\!-\!1.04)/140$.
An example of the results of the model fittings to the ASCA SIS 
spectra is shown in figure~\ref{fig:ascaRef}.
The best-fit parameters are listed in table~\ref{tbl:ascaRef}.
The time histories of the line flux and the continuum flux at 8--10~keV
for the ASCA data are plotted in figure~\ref{fig:ascaLCfe}.
No significant change of the line flux can be seen on time scales 
of $10^5$--$10^6$~s.
As can be seen from the table, the values of $N_{\rm H}^1$, $C_{\rm F}^1$
and $N_{\rm H}^2$ obtained from the spectral fits to the 6 sets of ASCA
spectra significantly change on time scales of $10^5$--$10^6$~s.
The relative amplitude of the variations of $N_{\rm H}^1$ and
$N_{\rm H}^2$ are $\sim$20\% and $\sim$12\%, respectively.

Finally, we investigated the parameters of the narrow line component.
We optimized the center energy and the intrinsic width of the narrow
line component in the ``reflection'' model by fitting it to the
average spectra of the ASCA observations in 2000.
The line center energy and the intrinsic width were found to be
$6.38\pm0.03$~keV and less than 92~eV (gaussian $\sigma$), respectively.

\subsection{Relation between the Line and the Continuum Fluxes}

An analysis of the time-resolved energy spectra showed that the iron line 
flux was consistent with being constant on time scales of $10^5$--$10^6$~s,
while significant variations were seen in its flux on time scales of
$10^6$--$10^7$~s.
In order to study the origin of the line emission, the relation between
the continuum flux and the line flux was studied.  
Here, the continuum flux was calculated in the energy band of 
8--10~keV, which is higher than the K edge energy of neutral iron.   
In this energy range, the flux variation should
reflect the variation of the intrinsic, power-law component.

Because the line emission probably results from a reprocessing of 
the continuum X-rays by matter ambient to the continuum source, 
the light curve of the line flux might suffer from some amount of 
smearing and/or time delay to the continuum flux light curve.
In fact, if we compare the light curves of the continuum flux and
the line flux (figure~\ref{fig:rxteFe}), the line flux variation
seems to follow the smeared variations of the continuum flux.
In order to see the effect of smearing, we applied smearing to the
continuum light curve using a following simple method.
We redistributed the continuum flux of $i$-th time bin into $N$ bins
starting from $i$-th bin.    
Thus, the smeared light curve is expressed as
\begin{equation}
G_i = \sum_{j=i-N+1}^{i} \frac{F_j}{N},
\end{equation}
where $F_i$ is the flux of $i$-th bin in the original light curve,
$G_i$ the flux of $i$-th bin in the smeared light curve, and $N$
the number of bins to smear.

\begin{figure}
\begin{center}
\FigureFile(85mm,60mm){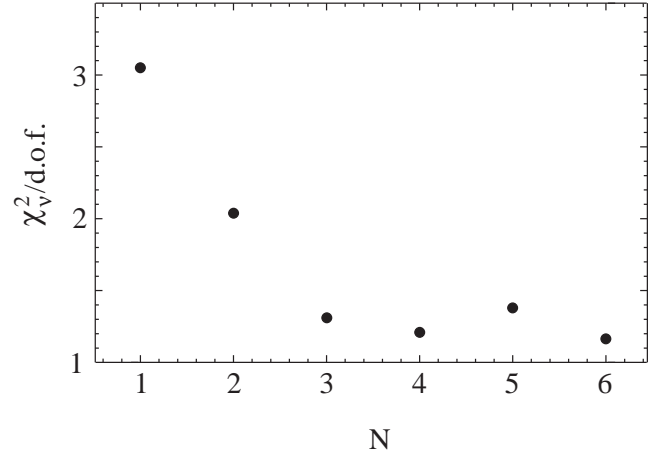}
\end{center}
\caption{Reduced $\chi^2$ value when we compare the smeared light
curve of the continuum flux with that of the line flux, as a function of a
parameter, $N$, representing the degree of smearing.  The reduced $\chi^2$
value becomes acceptable when $N$ increases up to 4.
} \label{fig:rxteSmear}
\end{figure}

\begin{figure}
\begin{center}
\FigureFile(85mm,60mm){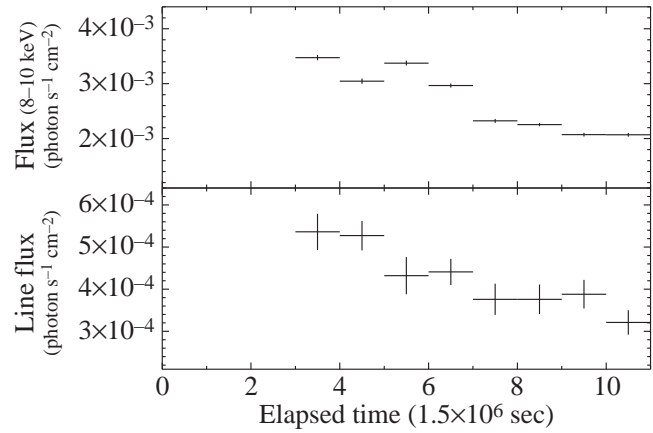}
\end{center}
\caption{Light curves of the line flux and the smeared continuum flux.
For the calculation method of smearing, see text.  
A good agreement can be seen between them.
} \label{fig:rxteSmComp}
\end{figure}

We then fitted the smeared light curve to the line flux light curve.
We increased $N$ from 1 until we obtain an acceptable fit.
Note that $N=1$ corresponds to the case without smearing.
The relation between $N$ and $\chi^2/{\rm d.o.f.}$ is plotted in
figure~\ref{fig:rxteSmear}.
We could obtain an acceptable agreement with $N=4$ between the smeared 
light curve of the continuum flux and the line flux light curve.
The smeared light curve in the case of $N=4$ is compared with the
light curve of the line flux in figure~\ref{fig:rxteSmComp}.
The above-mentioned smearing algorithm of the continuum flux also 
introduces a delay of $N/2$ bins together with smearing of $N/2$ bins
to both sides.     Thus the typical delay and smearing time scale
may be regarded as $3\times 10^6$~s.

We then checked whether or not smearing of $6\times 10^6$~s can explain
the absence of a correlation between the continuum flux and
the line flux on shorter time scales for the ASCA data.  Since the ASCA
data is shorter than $6\times 10^6$~s, we could not smear the continuum
flux light curve directly.   Hence, we assumed that the observed light curve
of the continuum flux in the ASCA observations repeats periodically.
We then smeared the assumed light curve using the method described above.
The fractional variation (i.e.\ standard deviation divided by the average)
of the continuum flux was originally $\sim$0.25, but was reduced to
$\sim$0.06 by smearing.
The reduced value is consistent with the observed upper limit, 7\%,
of the fractional variation of the line flux in the ASCA observations.

\section{Discussion}

We analyzed the ASCA data obtained in 2000 May which covered
time intervals of $1.1\times 10^6$~s.   The data were divided into 6
sets with an integration time of $1.8\times 10^5$~s each to study the
time variabilities of the spectrum.   In order to analyze the excess
component separately from the continuum, we masked a range of 4--8~keV
in spectral fitting and determined the continuum model.
We found that a power law with a photon index of $\sim$1.55 modified by
two absorbers with different column densities and covering fractions can
smoothly connect the two spectral parts at 2--4~keV and 8--10~keV\@.
Above this continuum, an excess is clearly detected in 4.5--7.5~keV, 
which has a broad and skewed feature.
The feature has a prominent narrow peak at 6.4~keV, but accompanies
a broad red wing at 4.5--6.0~keV and a blue wing at 6.8--7.5~keV\@.
The flux and shape of the excess feature obtained from each of the 6
spectra were compared with one another.
It is found that the excess flux is consistent with being stable,
concerning both the flux and the spectral shape, over the observations
of $1.1\times 10^6$~s.

We also analyzed data obtained by RXTE from January through July in
1999 in order to search for any time variations of the iron line 
on a time scale of $>\!10^6$~s.
We obtained 11 sets of time-resolved energy spectra,
each of which covers a time interval of $1.5\times 10^6$~s.
If we selected only the energy bands free from the iron structures,
i.e.\ the 2.8--4.0~keV and 8--10~keV bands, each energy spectrum could
be reproduced again by a power law with a photon index of $\sim$1.5
modified by the partial covering absorption.
This result is consistent with that of ASCA\@.
In 4.5--7.5~keV, an excess flux over the continuum is clearly seen as
a broad feature, in which significant flux variations are detected.
However, no clear change was noticed in its profile.
The absence of a profile change in spite of the significant flux variations
in the broad and skewed feature strongly indicates that the feature is
produced through a single mechanism.

We studied the time scale of the flux variations of the narrow line.
It was found that the flux variations are not significant on time scales
of $10^5$--$10^6$~s, whereas they become significant
on time scales of $10^6$--$10^7$~s.
We also studied whether or not an introduction of a smearing effect can
improve the correlation between the line flux and the continuum flux.
We redistributed the continuum flux light curve with a simple method
approximating a smear and delay.    We found that the agreement between
the line flux light curve and the smeared light curve of the continuum
flux becomes acceptable when we introduce a smear and a delay on a time
scale of $3\times 10^6$~s.
These results concerning the line flux variations and the effect 
of a smear strongly suggest that the line emitting region should 
have a size extent of 10$^{17}$~cm.

We further studied the time scale of the variation in the absorption column
density.   A column density of the order of $\sim\!10^{23}$ atom~cm$^{-2}$
is necessary to reproduce the continuum spectrum, and is found to vary
significantly on a time scale of $10^5$--$10^6$~s with a relative 
amplitude of $\sim$20\%.
Taking account of the relative amplitude of the variation, the size
of the absorber is indicated to be no larger than $10^{17}$~cm,
which is just the size of the line emitting region.
The absorber should be around, or inside, the line emitting region.

We tried a continuum plus a disk-line model to reproduce the energy
spectra, and found that this model can reproduce the energy spectra
at 2--10~keV for both ASCA and RXTE data.
As a result of the previous discussion, the size of the line emitting 
region should be as large as $10^{17}$~cm.
In the disk-line model, the line is assumed to be emitted from a region
with a size of several to ten times the Schwarzschild radius.
If the size of $10^{17}$~cm deduced from observational results 
corresponds to several to ten-times the Schwarzschild radius, 
we need to assume a central mass close to $10^{11}$~$M_\odot$.
This central mass is not consistent with $10^7$~$M_\odot$ estimated
for NGC~4151 from various methods, e.g.\ X-ray variability time scale 
\citep{hay98} and the analysis of the high ionization lines in the broad 
line region \citep{cla87,ulr96}.
This mass is rather close to the typical mass of a galaxy,
and would be too large as the mass of the central black hole.

The presence of the absorber around or inside the line emitting region would
also be difficult to explain in the context of the ``disk-line'' model.
If the line emitting region has a size of several to ten-times 
the Schwarzschild radius, the absorber should necessarily be 
located at a region very close to the central black hole.
This may not be consistent with the Unified scheme in which the 
heavy absorption as seen in NGC~4151 is considered to be due
to a dusty torus around the central active region.
We obtain an inclination angle of the disk to be $\sim\!20^\circ$.
This inclination angle is smaller than the generally accepted
value $63^\circ$, which is supported by the Chandra observation
\citep{ogl00} and the optical observations.
This is also a disadvantage of the ``disk-line'' model.

As discussed above, there exist some serious difficulties in the disk-line
model, which are not consistent with a reasonable, common picture of AGN,
although we cannot completely exclude its possibility.
We consider here an alternative explanation for the origin of the broad and
skewed profile, in which the narrow peak at 6.4~keV should be a narrow
line and the red and blue wings should be a part of the continuum.
However, according to the constant profile of the broad feature around
5--7~keV, the narrow line and the continuum should have a strong physical
connection.   The line energy at 6.4~keV strongly suggests that the line
is the fluorescent K-line from neutral or low ionization iron.
If the line is really emitted through a fluorescent process, it implies
that some amount of matter exists in the vicinity of the X-ray source.
This matter should be irradiated by X-rays from the X-ray source
and the fluorescent lines should be emitted there.
If this is the case, continuum X-rays should also be emitted from the matter
through electron scattering of irradiating X-rays.
This reflected component should be observed together with the fluorescent
line, and could be the red and blue wings.

When an X-ray from the source is absorbed by an iron atom in the X-ray
reflector through photo-ionization of a K-electron,
a fluorescent K-line is emitted with a certain probability called a
fluorescent yield.
On the other hand, when an X-ray from the source hits an electron in
the reflector, it could be re-emitted towards us through Thomson
(Compton) scattering.
The cross section of Thomson scattering is about $10^{-24}$~cm$^2$.
Hence, if the column density of the X-ray reflector is sufficiently large,
an X-ray penetrated into the reflector would, on average, experiences
photo-electric absorption by matter with a column density of about
$10^{24}$~cm$^{-2}$ before being scattered by an electron.
As a result, X-rays reflected by sufficiently thick matter should have
a spectrum with a continuum shape being the same as the intrinsic 
X-rays, but suffering from photo-electric absorption by matter with 
a column density of about $10^{24}$~cm$^{-2}$.
In fact, the broad and skewed feature can be reproduced by
a model consisting of a narrow line at 6.4~keV, and a power law
with the same slope as the remaining continuum and with a photo-electric
absorption by a column of $10^{24}$~cm$^{-2}$.
The equivalent width of the 6.4~keV line with respect to the heavily 
absorbed component is about 2~keV\@.  This is roughly consistent with 
a value expected from a case when the heavily absorbed component is 
just the reflected component (see e.g.\ \cite{ino85}).

If we adopt the above model for the broad and skewed feature around
4.5--8~keV, the total spectrum becomes to have three differently
absorbed components with the same continuum shape.
The absorption column densities are about $5\times10^{22}$~cm$^{-2}$,
(1--2)$\times 10^{23}$~cm$^{-2}$, and about $10^{24}$~cm$^{-2}$.
Among the three components, two components absorbed by smaller columns
are found to be variable on a time scale of $10^5$--$10^6$~s,
whereas the component with the absorption by $\sim\! 10^{24}$~cm$^{-2}$
was steady on that time scale.
This strongly suggests that the heavily absorbed component has a different
origin from the other two components.
The time variability of the heavily absorbed component is rather similar
to that of the fluorescent iron line.
This indicates a strong physical coupling between the fluorescent iron line
and the heavily absorbed component, and strongly supports the idea that the
heavily absorbed component is a reflected emission by cold matter which
is also responsible for the fluorescent iron line.

The above model fits both the ASCA spectra at 2--10~keV
and the RXTE spectra at 2--24~keV well.
The model can also fit the data in a higher energy range.
Actually, if we extrapolate the best-fit model in the energy range
of 2.8--24~keV to 100~keV, no significant discrepancy is recognized
between the model and the observed energy spectra.
Therefore, this model can explain wider range of energy spectra
without the spectral break than the model based on the disk-line.
If the reflection is really at work, the covering fraction of the
absorber may be related to the solid angle of the reflector, $\Omega$,
subtending to the X-ray source.
The covering fraction that we obtained for an absorber of
$10^{24}$~cm$^{-2}$ is $\sim\!0.47$, which corresponds to
$\Omega \sim 0.47\times4\pi$.
The reflection structure was detected in the energy spectra of
NGC~4151 with the Ginga/OSSE data \citep{zdz96} and 
the BeppoSAX data \citep{pir98}.
The BeppoSAX data indicated that the relative contribution of the 
reflection changed with time.
The covering fraction of the reflector so far reported ranges 
over $\Omega \sim (0.0\!-\!0.4)\times4\pi$ \citep{zdz96,pir98}.
The covering fraction which we obtained is slightly larger than this
range, but is not very different, if we consider the different 
modeling of the reflection component.

Based on the reflection model, we may be able to constrain the mass
of the central black hole using the iron line parameters.
 From the model fitting, we found that the iron line is consistent
with having no intrinsic width, and its upper limit is 92 eV\@.
We also found that the iron line emitting region has a size of $10^{17}$~cm.
It may be natural to assume that the line emission region is located
around the central black hole and is rotating at the Kepler velocity
around the black hole.
Then, the line width may be determined by the Doppler effect;
the upper limit of the line width can be related to the upper limit
of the Kepler velocity as $v/c \sim 0.014$.
This may be converted to the mass of the black hole using the relation
$v^2=GM/r$, which yields the upper limit of the black hole mass,
$\sim\! 10^8$~$M_{\odot}$.    Here, we assume $r \sim 10^{17}$~cm.
This is consistent with the estimated mass of $10^7$~$M_{\odot}$.

If the mass of the central black hole is $10^7$~$M_{\odot}$,
a few times 10$^{16}$~cm corresponds to $10^4$~$r_{\rm Sch}$.
At this distance, the X-ray reflector should exist.
Since the reflected component shows evidence of absorption by
a column of $10^{24}$~cm$^{-2}$, the reflector should be Thomson thick.
The solid angle of the reflector, as viewed from the central X-ray 
source, should be about $0.47 \times 4\pi$.
This reflector would be a so-called dust torus, which is generally believed
to exist around the central engine in Seyfert galaxies.

It has been shown that the continuum spectrum of NGC~4151 needs
a partial covering by X-ray absorbers with a column density of
(1--2)$\times 10^{23}$~cm$^{-2}$.
The absorption column and the covering factor have been found to vary on
a time scale of $10^5$--$10^6$~s.
As already discussed above, this leaky absorber should exist around or
inside the line emitting region.
If a dusty torus emits the line, a geometrical relation between a
dusty torus and a broad line region seems to be consistent with the
observed constraint on the geometrical relation between the line emitting
region and the leaky absorber.
The absorbers could be relatively cold clouds in the broad line region.
If so, the typical time scale of a change of the leaky absorption should
be roughly estimated by $R/V$, where $R$ is the size of the X-ray emitting
region and $V$ the velocity of the X-ray absorbing clouds.
Since the typical velocity of the broad line clouds is several times
$10^8$~cm~s$^{-1}$ and the typical time scale of the absorption change
is a few times $10^5$~s, the size of the emission
region may be about $10^{14}$~cm.
This size corresponds to a few ten-times $r_{\rm Sch}$ of a black hole
with a mass of $10^7$~$M_{\odot}$, and is consistent with a natural
expectation that a region close to the central black hole should be
emitting the continuum X-rays.

 From these results and considerations, we conclude that the disk line
model has difficulty to explain the spectral variations of NGC~4151,
and the reflection model is much more plausible.

\clearpage

% Table 1
\begin{table}[htb]
\caption{Selection criteria of ASCA data.}\label{tbl:crit}
\begin{center}
\begin{tabular}{ll}\hline\hline
   Telemetry bit rate   \dotfill              & High and Medium        \\
   Pointing offset from the target \dotfill & $\leq 0.05^\circ$   \\
   Elevation angle from the Earth limb   \dotfill & $\geq 5^\circ$      \\
   Cut-off rigidity                \dotfill & $\geq $ 6 GV           \\
   Elevation angle from the day Earth (SIS only) \dotfill & $\geq 25^\circ$ \\
\hline\hline
\end{tabular}
\end{center}
\end{table}

% Table 2
%
\begin{table}[htb]
\caption{Best-fit parameters of the ``disk-line'' model for the 6
energy spectra from ASCA.}
\label{tbl:ascaPar}
\begin{center}
\begin{tabular}{lcccccc}\hline\hline
Sequence number & 1 & 2 & 3 & 4 & 5 & 6 \\\hline
$N_{\rm H}^1$ [$10^{22}$ atom cm$^{-2}$] \dotfill &
         $11\pm2$  & $9\pm3$ & $7\pm2$ & $5\pm3$ & $7\pm3$  & $7\pm3$ \\
$C_{\rm F}^1$ [\%] \dotfill &
         $87\pm4$  & $90\pm4$ & $90\pm4$  & $93\pm4$  & $87\pm5$ & $84\pm5$ \\
Inclination angle [deg] \dotfill &
         $21\pm5$  & $18\pm4$  & $21\pm5$  & $22\pm6$ & $21\pm6$ & $19\pm6$ \\
line flux [$10^{-4}$ photon cm$^{-2}$ s$^{-1}$] \dotfill &
         $3.0\pm0.3$  & $3.0\pm0.3$  & $3.0\pm0.3$  & $3.5\pm0.3$ &
         $2.8\pm0.3$  & $3.0\pm0.3$ \\
$\chi^2_\nu/{\rm d.o.f.}$ \dotfill &
         1.10/140 & 1.02/140  & 0.99/140 & 1.03/140 & 1.02/140 & 1.03/140 \\
\hline\hline
\multicolumn{7}{l}{Note: See equation~(\ref{eq:dualAbs}) for the continuum
model.   Errors are at the 90\% confidence limit.}\\
\end{tabular}
\end{center}
\end{table}

% Table 3
%
\begin{table}[htb]
\caption{Best-fit parameters of the ``disk-line'' model for the 11
energy spectra from RXTE PCA.}
\label{tbl:rxtePar}
\begin{center}
\begin{tabular}{lcccccc}\hline\hline
Sequence number  \dotfill  & 1 & 2 & 3 & 4 & 5 & 6  \\  \hline
Photon index ($\Gamma_1$)  \dotfill & $1.50\pm0.11$  & $1.45\pm0.13$
         & $1.58\pm0.11$  & $1.60\pm0.11$  & 1.55$^{+0.14}_{-0.12}$
         & 1.45$^{+0.10}_{-0.12}$   \\
Break energy [keV]  \dotfill  & $17\pm2$  & $18\pm2$  & $17\pm2$  & $18\pm2$
         & $16\pm3$  & $19\pm3$ \\
Photon index ($\Gamma_2$)  \dotfill  & $1.92\pm0.22$  & $2.02^{+0.20}_{-0.18}$
         & $1.89\pm0.21$  & $2.01\pm0.23$  & 1.89$^{+0.18}_{-0.22}$
         & 1.87$^{+0.24}_{-0.26}$ \\
$N_{\rm H}^{1}$ [10$^{22}$ atom cm$^{-2}$] \dotfill  & $7.9^{+1.9}_{-2.1}$
         & $4.4^{+2.0}_{-1.8}$  & $4.9\pm2.2$  & $6.1^{+2.0}_{-1.8}$
         & $3.4\pm2.1$  & $3.8\pm2.2$   \\
$C_{\rm F}^1$ [\%] \dotfill  & $82\pm15$  & $\geq70$
         & $83\pm17$  & $75\pm18$  & $\geq68$ & $\geq79$   \\
Line flux [10$^{-4}$ photon cm$^{-2}$ s$^{-1}$] \dotfill
         & $8.9\pm0.6$  & $9.1\pm0.6$  & $9.8\pm0.6$  & $9.4\pm0.6$
         & $9.0\pm0.6$  & $8.0\pm0.6$    \\
$\chi^2_\nu/{\rm d.o.f.}$ \dotfill  & 1.07/46  & 1.40/46  & 1.12/46  & 1.48/46
         & 1.14/46  & 1.27/46  \\ \hline\hline
Sequence number  \dotfill  & 7 & 8 & 9 & 10 & 11   \\  \hline
Photon index ($\Gamma_1$)  \dotfill &  $1.51\pm0.12$  & $1.51^{+0.15}_{-0.13}$
         & $1.50\pm0.14$  & $1.45\pm0.13$ & $1.45\pm0.11$  \\
Break energy [keV]  \dotfill  & $17\pm2$ & $16\pm2$ & $17\pm3$
         & $19\pm2$  &  $16\pm2$  \\
Photon index ($\Gamma_2$)  \dotfill  & $1.92^{+0.21}_{-0.24}$
         & $1.87^{+0.26}_{-0.28}$  & $1.89^{+0.26}_{-0.24}$
         & $2.01^{+0.23}_{-0.26}$  & $1.82^{+0.26}_{-0.24}$ \\
$N_{\rm H}^{1}$ [10$^{22}$ atom cm$^{-2}$] \dotfill  & $5.3\pm2.3$
         & $7.4^{+2.5}_{-2.7}$ & 9.7$^{+2.3}_{-2.1}$ & $7.7^{+2.0}_{-2.4}$
         & $4.5\pm2.3$ \\
$C_{\rm F}^1$ [\%] \dotfill  &  $87\pm12$  &  $85^{+14}_{-16}$
         & $80^{+17}_{-15}$ & $85^{+13}_{-16}$ & $\geq 78$ \\
Line flux [10$^{-4}$ photon cm$^{-2}$ s$^{-1}$] \dotfill
         &  $7.5\pm0.6$  &  $6.5\pm0.6$  & $6.0\pm0.6$  & $6.7\pm0.6$
         & $6.1\pm0.5$ \\
$\chi^2_\nu/{\rm d.o.f.}$ \dotfill  & 1.12/46 & 0.99/46 & 1.39/46 & 1.28/46
         & 1.20/46 \\ \hline\hline
\multicolumn{7}{l}{Note: See equation~(\ref{eq:dualAbs}) for the continuum
model.   Errors are at the 90\% confidence limit.}\\
\end{tabular}
\end{center}
\end{table}

% Table.4 : The best fit parameters of the ``reflection'' model to the 11
%           PCA spectra.
\begin{table}[htb]
\caption{Best-fit parameters of the ``reflection'' model for the 11
energy spectra from RXTE PCA.}
\label{tbl:rxteRef}
\begin{center}
\begin{tabular}{lcccccc}\hline\hline
Sequence number  \dotfill  & 1 & 2 & 3 & 4 & 5 & 6  \\  \hline
Photon index \dotfill  & $1.90\pm0.12$  & $1.91^{+0.11}_{-0.13}$
         & $1.97\pm0.10$ & $2.00^{+0.11}_{-0.13}$ & $1.87\pm0.15$
         & $1.89\pm0.10$ \\
$N_{\rm H}^1$ [10$^{22}$ atom cm$^{-2}$] \dotfill  & $12\pm2$
         & $11\pm2$ & $10\pm3$  &  $14\pm2$ & $10\pm3$ & $12\pm2$ \\
$C_{\rm{F}}^1$ [\%] \dotfill   &  $84\pm6$  &  $82\pm4$  & $77\pm6$
         & $77\pm4$ &  $78\pm8$  & $75\pm7$ \\
$N_{\rm H}^3$ [10$^{22}$ atom cm$^{-2}$] \dotfill  & $106^{+25}_{-27}$
         & $109\pm27$  & $99^{+25}_{-22}$  & $106^{+22}_{-22}$
         & $90\pm30$  & $117^{+29}_{-27}$ \\
$C_{\rm F}^2$ [\%] \dotfill   &  $53\pm13$  & $56\pm12$  & $52\pm11$
         & $61\pm13$ & $53\pm18$ & $37\pm10$ \\
Line flux [$10^{-4}$ photon cm$^{-2}$ s$^{-1}$] \dotfill  &  $4.9\pm0.6$
         & $5.1\pm0.6$ & $5.7\pm0.7$  & $5.4\pm0.6$ & $5.3\pm0.7$
         & $4.3\pm0.7$ \\
$\chi^2_\nu/{\rm d.o.f.}$ \dotfill  & 0.99/46 & 0.94/46 & 0.91/46 & 0.88/46
         & 0.97/46  &  0.87/46  \\ \hline\hline
Sequence number  \dotfill  & 7 & 8 & 9 & 10 & 11  \\  \hline
Photon index \dotfill  & $1.89^{+0.10}_{-0.12}$ & $1.82^{+0.15}_{-0.13}$
         & $1.85\pm0.14$ & $1.89\pm0.13$ & $1.83\pm0.12$ \\
$N_{\rm H}^1$ [10$^{22}$ atom cm$^{-2}$] \dotfill  & $12\pm1$ & $14\pm2$
         & $13\pm2$ & $18\pm2$ & $12\pm1$ \\
$C_{\rm{F}}^1$ [\%] \dotfill   & $76\pm5$ & $81\pm6$ & $83\pm8$ & $85\pm4$
         & $82\pm6$ \\
$N_{\rm H}^3$ [10$^{22}$ atom cm$^{-2}$] \dotfill  & $120\pm34$
         & $110\pm35$  & $90\pm33$  & $100\pm34$ & $100^{+26}_{-37}$ \\
$C_{\rm F}^2$ [\%] \dotfill   & $35\pm10$ & $33\pm11$ & $43^{+21}_{-19}$
         & $66\pm13$  & $45\pm11$ \\
Line flux [10$^{-4}$ photon cm$^{-2}$ s$^{-1}$] \dotfill  & $4.4\pm0.6$
         & $3.8\pm0.6$ & $3.8\pm0.6$ & $3.9\pm0.6$ & $3.2\pm0.5$ \\
$\chi^2_\nu/{\rm d.o.f.}$ \dotfill  & 0.98/46 & 0.91/46 & 0.99/46 & 0.91/46
         &  0.90/46  \\\hline\hline
\multicolumn{7}{l}{Note: Errors are at the 90\% confidence limit.}\\
\end{tabular}
\end{center}
\end{table}

% Table.5 : The best fit parameters of the ``reflection'' model to the 6
%           ASCA spectra.
\begin{table}[htb]
\caption{Best-fit parameters of the ``reflection'' model for the 6
energy spectra from ASCA.}
\label{tbl:ascaRef}
\begin{center}
\begin{tabular}{lcccccc}\hline\hline
Sequence number  \dotfill  & 1 & 2 & 3 & 4 & 5 & 6  \\  \hline
$N_{\rm H}^1$ [$10^{22}$ atom cm$^{-2}$] \dotfill   & $17\pm2$ & $15\pm3$
         & $16\pm3$ & $12\pm3$ & $18\pm3$ & $19\pm3$  \\
$C_{\rm F}^1$ [\%] \dotfill  &  $67\pm5$  &  $59\pm7$  &  $50\pm7$
         & $46\pm5$ & $47\pm5$ & $48\pm5$ \\
$N_{\rm H}^2$ [$10^{22}$ atom cm$^{-2}$] \dotfill  & $4.6\pm0.4$
         & $5.3\pm0.5$  & $5.6\pm0.5$  & $4.9\pm0.6$ & $4.9\pm0.5$
         & $4.6\pm0.5$ \\
line flux [$10^{-4}$ photon cm$^{-2}$ s$^{-1}$] \dotfill  & $1.9\pm0.2$
         & $1.9\pm0.3$  & $1.8\pm0.3$  & $1.9\pm0.3$ & $1.8\pm0.3$
         & $2.0\pm0.3$  \\
$\chi^2_\nu/{\rm d.o.f.}$  \dotfill  & 1.04/140  & 0.97/140 & 0.94/140 
	& 0.99/140 & 0.99/140  & 1.02/140 \\\hline\hline
\multicolumn{7}{p{15cm}}{Note: The model of equation~(\ref{eq:ref}) is used.
The photon index, the column density of $N_{\rm H}^3$, and the covering
fraction of $C_{\rm F}^2$ are fixed to 1.9, $10^{24}$ atom~cm$^{-2}$, 
and 47\%, respectively.    Errors are at the 90\% confidence limit.}
\end{tabular}
\end{center}
\end{table}

\clearpage

% Fig. 1
%

% Fig.2 : The average spectra of GIS and SIS

% Fig.3 : Spectral ratios of 6 ASCA spectra to the average

% Fig.4 : Disk-line feature of the average spectrum

% Fig.5 : Residuals in 4.0 -- 8.0 keV of the six SIS spectra.

% Fig.6 : PHA ratio of the excess feature to the average

% Fig.7 :

% Fig.8 : PCA spectrum of NGC~4151 averaged over the RXTE observations.

% Fig.9 : Residuals of the 11 spectra

% Fig.10 : Ratios of the 11 residual spectra to their average

% Fig.11 :

% Fig.12 : The variations of the line flux and the continuum flux in the
%          RXTE observations.

% Fig.13 :

% Fig.14 :

% Fig.15

% Fig.16

\end{document}